# OpenFPM: A scalable open framework for particle and particle-mesh codes on parallel computers

Pietro Incardona[a,b], Antonio Leo[a], Yaroslav Zaluzhnyi[a], Rajesh Ramaswamy[d], Ivo F. Sbalzarini[a,b,c,*]

[a]*Chair of Scientific Computing for Systems Biology, Faculty of Computer Science, TU Dresden.*
[b]*MOSAIC Group, Center for Systems Biology Dresden.*
[c]*Max Planck Institute of Molecular Cell Biology and Genetics, Dresden.*
[d]*Max Planck Institute for the Physics of Complex Systems, Dresden. (now at: EMD Serono, Inc., Boston.)*

**Abstract**

Scalable and efficient numerical simulations continue to gain importance, as computation is firmly established as the third pillar of discovery, alongside theory and experiment. Meanwhile, the performance of computing hardware grows through increasing heterogeneous parallelism, enabling simulations of ever more complex models. However, efficiently implementing scalable codes on heterogeneous, distributed hardware systems becomes the bottleneck. This bottleneck can be alleviated by intermediate software layers that provide higher-level abstractions closer to the problem domain, hence allowing the computational scientist to focus on the simulation. Here, we present OpenFPM, an open and scalable framework that provides an abstraction layer for numerical simulations using particles and/or meshes. OpenFPM provides transparent and scalable infrastructure for shared-memory and distributed-memory implementations of particles-only and hybrid particle-mesh simulations of both discrete and continuous models, as well as non-simulation codes. This infrastructure is complemented with portable implementations of frequently used numerical routines, as well as interfaces to third-party libraries. We present the architecture and design of OpenFPM, detail the underlying abstractions, and benchmark the framework in applications ranging from Smoothed-Particle Hydrodynamics (SPH) to Molecular Dynamics (MD), Discrete Element Methods (DEM), Vortex Methods, stencil codes, high-dimensional Monte Carlo sampling (CMA-ES), and Reaction-Diffusion solvers, comparing it to the current state of the art and existing software frameworks.

## 1. Introduction

Computer simulations enable the study of complex models beyond what is possible analytically. The primary goal of a simulation therefore is the scientific or engineering result obtained from the computer experiment. The majority of time, however, is usually spent developing, testing, and optimizing the simulation codes. This is mainly due to the "knowledge gap", stating that the increasingly specialized knowledge required to efficiently use modern heterogeneous supercomputing platforms is found in only a small group of people [1]. The knowledge gap can be successfully addressed by abstracting the algorithmic implementation from the computer system platform.

Such abstraction has a long tradition in computational science. It is typically provided as *programming language extensions*, *high-level programming languages*, *software libraries*, or as a *framework* combining programming language extensions and libraries. For high-performance parallel computing, programming language extensions include OpenACC [2] and OpenMP [3] that provide a directive-based parallel programming model, CUDA [4] and OpenCL [5] for GPGPU and accelerator programming, co-array Fortran (CAF) [6], High-Performance Fortran (HPF) [7], and Unified Parallel C (UPC) [8]. Examples of high-level programming languages for parallel computing include Linda [9], providing a model for coordination and communication between parallel processes, Vectoral [10] for direct vector-processor programming, and Julia [11] designed for high-performance numerical analysis. Examples of software libraries for parallel HPC include implementations of the Message Passing Interface (MPI) [12] standard like OpenMPI [13] and MPICH [14], HPX [15] as a runtime system for parallel and distributed applications, the DASH implementation of the PGAS model providing general-purpose distributed data-structures [16], and Charm++ [17] as an example of a framework for distributed parallel programming. Beyond HPC, a number of languages, libraries, and problem-solving environments [18] for simulation exist that focus on sequential processing, including the equation-based simulation language Modelica [19], a Matlab-based compiler [20], and the scientific computing environment FALCON [20].

Such languages and libraries can greatly reduce code-development overhead and render hardware platforms more accessible to a wide user base. However, the abstractions on which these libraries and languages are based cannot be universal and concise at the same time. Fine abstractions, like those in MPI [21] can be universal (i.e., every parallel algorithm can be implemented with MPI), but may remain hard to use. Coarse abstractions, like the parMETIS library for parallel graph partitioning [22], are easy to use, but provide only limited flexibility (i.e., parMETIS cannot do parallel FFTs). The knowledge gap reduction that is achievable with general-purpose abstractions, like the ones above, is therefore limited by the trade-off between generality and ease of use.

Further reduction of the knowledge gap is possible using abstractions that are specific to a certain application domain. This has been successfully exploited by numerical simulation libraries like OpenFOAM for finite-volume simulations [23], DUNE [24] and Trilinos [25] for finite-element simulations, DualSPHysics [26] for Smoothed-Particle Hydrodynamics (SPH) simulations, NAMD [27] and LAMMPS [28] for Molecular Dynamics (MD) and Dissipative Particle Dynamics (DPD) simulations, LibGeoDecomp [29] for cell-based decomposition codes, and AMReX [30] for Adaptive Mesh Refinement (AMR). Examples of domain-specific programming languages for parallel numerical simulations include DOLFIN [31], the programing language of the FEniCS framework for finite-element simulations [32], and Liszt [33], a domain-specific language for mesh stencil codes. Among numerical simulation frameworks, particle methods are particularly appealing from a software-engineering viewpoint, because they can be used to simulate models of all four kinds: discrete, continuous, deterministic, stochastic. When used to simulate discrete models, particles naturally correspond to the modeled entities, e.g. atoms in molecular dynamics or cars in road traffic simulation. When simulating continuous models, particles correspond to (Lagrangian) tracer points or mathematical collocation points, e.g., as in SPH [34] or Particle Strength Exchange (PSE) [35, 36, 37] simulations. Naturally, particle interactions and dynamics can be either deterministic or stochastic. For particle methods, mainly two frameworks for distributed parallel computing exist: POOMA [38] and the Parallel Particle Mesh (PPM) library [39, 40] with its domain-specific Parallel Particle Mesh Language (PPML) [41, 42]. While both have successfully provided abstractions for rapid development of scalable parallel implementations of particle and particle-mesh methods, they seem to be discontinued. However, all these languages and libraries have successfully demonstrated the benefits of domain-specific abstraction, and helped close the knowledge gap in scientific high-performance computing.

Based on this past success, and motivated by the absence of actively maintained frameworks for parallel particle-mesh simulations, we here present OpenFPM, an open-source C++ framework for parallel particles-only and hybrid particle-mesh codes. OpenFPM is intended as a successor to the discontinued PPM Library [39, 40] using advanced methods from scientific software engineering, such as template meta-programming (TMP). OpenFPM combines the most general formulation of particle methods with classic mesh-based approaches. A particle is defined as a point in an arbitrary-dimensional space that carries an arbitrary number of arbitrary data structures. A mesh is a regular division of the space into polyhedra (Cartesian at the time of writing) with support for local refinement. OpenFPM uses C++ TMP to transparently handle simulation domains of any dimension, and to allow particles to carry any C++ object as a property, including objects from user-defined classes. OpenFPM also provides its own memory manager and runtime dynamic load-balancing in order to transparently distribute the data and the work among the processors, and to dynamically adapt to changes in local mesh resolution or particle density during a simulation. The functionality of OpenFPM therefore goes beyond that of the PPM Library, and it relaxes PPM's most salient limitations, including the limitation to 2D and 3D simulations and the limitation to primitive types as particle properties. It also extends the capabilities by adding transparent dynamic load balancing, support for accelerator hardware, and automatic memory layout optimization. Using OpenFPM is further facilitated by a complete and up-to-date documentation, as well as a series of tutorial videos and example codes. It is actively supported in the long term with new functionality continuously being added. OpenFPM implements the same abstractions as the PPM Library [1], rendering it easy to understand for experienced PPM users and for novice developers alike.

## 2. Particle Methods

Particle methods provide a unifying algorithmic framework for simulating both discrete and continuous models. When simulating discrete models, particles directly represent the modeled entities. When simulating continuous models or numerically solving partial differential equations, particles represent mathematical collocation points. In both cases, a particle $p$ is described by its position $x_p \in \mathbb{R}^n$ and properties $w_{i,p}$ where different properties $i$ can be of different data type. Particles can do two things: they can interact with each other, and they can evolve. Interactions are limited to be pairwise. Three-particle and higher-order interactions are emulated as sequences of pairwise interactions. In addition, interactions are restricted to be additive, which ensures that the overall result is independent of the particle indexing order. After all interactions are computed, particles evolve their positions and properties according to pre-defined rules. These rules can be directly given by the model, such as update rules in a cellular automaton, or they may result from discretization of continuous differential operators, such as in time-integration schemes.

In the most general case, where each particle interacts with every other particle, the positions and properties hence evolve as:

$$\frac{\mathrm{d}\boldsymbol{x}_p}{\mathrm{d}t} = \sum_{q=1}^{N(t)} \boldsymbol{K}(\boldsymbol{x}_p, \boldsymbol{x}_q, \boldsymbol{w}_{i,p}, \boldsymbol{w}_q)$$

$$\frac{\mathrm{d}\boldsymbol{w}_p}{\mathrm{d}t} = \sum_{q=1}^{N(t)} \boldsymbol{F}(\boldsymbol{x}_p, \boldsymbol{x}_q, \boldsymbol{w}_{i,p}, \boldsymbol{w}_q)$$

with the total number of particles $N(t)$ at time $t$, and the interaction kernels $\boldsymbol{K}$ and $\boldsymbol{F}$ that encode the specific evolution rules of the model being simulated or the numerical method used. An important aspect of particle methods is that the total number of particles can adaptively change during a simulation. It is possible to locally add particles or remove them. This may represent discrete model dynamics, or can be used to implement adaptive-resolution solvers for continuous models.

In the above equations, the evolution of each particle depends on interactions with all $N$ (other) particles. This leads to a nominal computational cost that scales as $O(N^2)$ with more efficient approximation algorithms available [43, 44]. In many applications, however, the kernels $\boldsymbol{K}$ and $\boldsymbol{F}$ are local or compact, such that particles only need to interact within a finite radius. Together with efficient data structures, such as Cell Lists [45] or Verlet Lists [46], this reduces the computational cost to $O(N)$ *on average*. In other cases, the kernels can be decomposed into short- and long-range components and only the short-range component is directly evaluated on the particles, whereas the long-range interactions are computed on a uniform Cartesian background mesh [47]. Examples of such hybrid particle-mesh methods include the Ewald method for including electrostatic interactions in MD simulations [48] and remeshed vortex methods for solving the incompressible Navier-Stokes equations [49]. Hybrid particle-mesh methods allow each computational step to be performed in the better-suited formulation. When simulating continuous models, moment-conserving particle-mesh and mesh-particle interpolation are used to translate between the two discretizations [50].

## 3. The OpenFPM Library

OpenFPM is a portable open-source software library to implement scalable particle and hybrid particle-mesh simulations on distributed-memory parallel computer systems. OpenFPM is written entirely in C++ and provides the same abstractions as the PPM Library [39, 40]. These abstractions have been designed to be as coarse-grained as possible while still separating computation from communication [1]. OpenFPM provides portable and scalable infrastructure to implement custom particle and particle-mesh codes. It provides methods for domain decomposition, dynamic load balancing, communication abstractions, transparent handling of cross-processor interactions, checkpoint/restart facilities, as well as iterators for particles and meshes. This infrastructure is complemented by a set of frequently used numerical solvers, including an interface to the solvers in PetSc [51] and in Eigen [52].

OpenFPM provides flexibility through templated parametric data structures. A parameter in a data structure can, e.g., define the dimensionality of the space or the floating-point precision used. It can also add or modify the code of the data structure implementation, e.g., in order to change the internal memory layout. This enables hardware-targeting of data structures at compile time. In addition, OpenFPM enforces strong unit testing with strict code quality assurance. Its design follows modern programming techniques like Template Meta Programming (TMP) and the familiar configure/make installation process. OpenFPM release cycles follow the software engineering method of Continuous Integration and include static code analysis, automated performance testing, memory coherence and memory leak testing, code coverage analysis, and tested installation support for multiple operating systems and compilers. The full documentation of OpenFPM is automatically updated after every change of the source code. Updates are deployed online as source code, binaries for Windows (Cygwin), macOS, and Linux, as virtual-machine disk images with complete pre-installed OpenFPM, and as Docker containers for virtualized environments. This ensures code quality, simplifies installation, and renders the library fully open and accessible.

*3.1. OpenFPM abstractions*

OpenFPM provides transparently distributed data structures for particles and meshes in $n$-dimensional computational domains. Particle sets in OpenFPM are objects storing the positions $x_p$ and properties $w_{i,p}$ of all particles $p$ in the set. Multiple particle sets can be used concurrently. At the time of writing, meshes are implemented as regular Cartesian grids that do not require storing the position of each mesh node. This defines the data abstractions of OpenFPM: particle sets and meshes.

These data abstractions are distributed across across multiple computers or memory address spaces in a way that is transparent to the user. This is done by decomposing the $n$-dimensional simulation domain into sub-domains that are then assigned processors (Fig. 1). Each processor only stores the particles and mesh nodes inside its assigned subdomains. Each processor also only computes the interactions and values of its particles and mesh nodes, hence parallelizing both data and work. In order for all computations to be local, sub-domains are extended by a *ghost layer* (halo layer) around inter-processor boundaries (Fig. 1). The width of the ghost layers is given by the particle interaction radius or the radius of the mesh stencil.

Inter-processor communication in OpenFPM is done by communication abstractions called *Mappings*. Mappings only communicate, but do not compute. Different types of mappings are provided to distribute particle and/or mesh data according to a certain domain decomposition and to manage ghost layers. Once the ghost layers are populated, all computation can be done locally, hence cleanly separating communication and computation. This makes the communication overhead explicit to the user and enables optimization and predictive performance modeling.

*3.2. Domain decomposition*

The domain decomposition process in OpenFPM is divided into three phases: *decomposition*, *distribution*, and *sub-domain creation*. The decomposition stage divides the physical domain into Cartesian sub-sub-domains (small squares in Fig. 1). The number of sub-sub-domains generated is at least as large as the number of processors, but typically much larger.

In the second phase, the sub-sub-domains are assigned to processors (colors in Fig. 1). Because their number is larger than the number of processors, there is no trivial assignment. Instead, the additional degree of freedom can be used to improve load balance and to reduce communication overhead. An optimal mapping would ensure that each processor receives the same amount of computational work (in terms of wall-clock time), while the total volume of inter-processor communication is minimized. This problem can be modeled as a graph-partitioning problem where each sub-sub-domain is a vertex of the graph, and an undirected edge between two vertices indicates an exchange of data between the respective sub-sub-domains through their overlapping ghost layers. In order to account for varying particle density and different sub-sub-domain sizes, we assign to each vertex a computational cost $c_i$ that is proportional to the total amount of compute operations required in sub-sub-domain $i$. We also weight the edges $e_{i,j}$ between sub-sub-domains $i$ and $j$ proportional to the amount of data that needs to be exchanged between them. In an optimal assignment, the sum of the vertex weights is the same for all processors, while the sum of the weights of all edges crossing a processor boundary is minimal. While the optimal solution cannot efficiently be computed, several publicly available libraries compute approximate sub-optimal solutions to this problem. These libraries include Scotch [53], ParMetis [22], and Zoltan [54]. OpenFPM uses ParMetis because of the efficiency of the implemented algorithms. The approximate solution to the graph-partitioning problem computed by ParMetis then defines the assignment of sub-sub-domains to processors. Alternatively, the user can choose to distribute sub-sub-domains along a Hilbert-type space-filling curve.

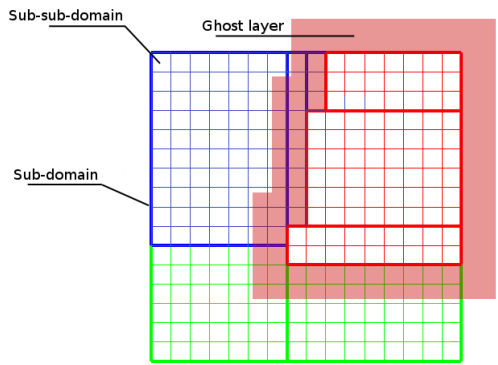

Figure 1: Domain decomposition in OpenFPM. The computational domain is decomposed into Cartesian sub-sub-domains (small squares) that are subsequently assigned to processors (colors). After assignment, cuboidal blocks of sub-sub-domains are merged to form the larger sub-domains (bold lines). One processor may have more than one sub-domain. Sub-domain borders at processor boundaries are extended by a ghost layer (shaded area, shown exemplarily for the red processor) in order to locally provide all data required for computations.

In the third step, the sub-sub-domains assigned to a processor are merged into larger sub-domains (bold lines in Fig. 1) wherever possible, in order to reduce the total ghost-layer volume, since sub-sub-domains on the same processor do not require a ghost layer between them. While this is natural for particles, the merging is important for meshes. The goal of sub-domain creation therefore is to merge sub-sub-domains such that the minimum number of sub-domains with smallest surface-to-volume ratio is created on each processor.

The sub-domain creation algorithm runs in parallel on each processor and starts from the first (by indexing order) sub-sub-domain on that processor, which is used as a seed. It then extends its boundaries uniformly in all directions. For example, in 2D, the box is enlarged by shifting the border by one sub-sub-domain in direction +X, +Y, -X, -Y. This procedure is iterated until the sub-domain border reaches an inter-processor boundary, or it is not possible to merge any more sub-sub-domains in any direction. A sub-domain is then created from all merged sub-sub-domains. The process then chooses the next (by indexing order) unassigned sub-sub-domain at the border of the just-created sub-domain and repeats until all sub-sub-domains have been assigned. Despite not finding the optimal solution, i.e., the one with the smallest number of

```
vector_dist<2,float, aggregate<float,                    (a)
                        float[3],
                        Point<3,double>,
                        openfpm::vector<float>>
```

```
vector_dist<2,float, aggregate<float,                    (b)
                        float[3],
                        Point<3,double>,
                        openfpm::vector<float>>,
            memory_traits_lin,
            CartDecomposition<2,float>,HeapMemory,
                        ParMetisDistribution<2,float>
            HeapMemory
            >
```

Figure 2: Data structures in OpenFPM are parametric, which allows to target them to different space dimensionality, data types, and hardware platforms at compile time. An example is shown for a particle set defined as a distributed vector. This includes mandatory parameters (a), such as the space dimension and the data types of the particle properties, but may also include optional parameters (b), such as the internal memory layout or the type of domain decomposition used for this data strcuture. This renders all data structures generic and independent of their actual implementation, with optimized internal code automatically generated by the compiler.

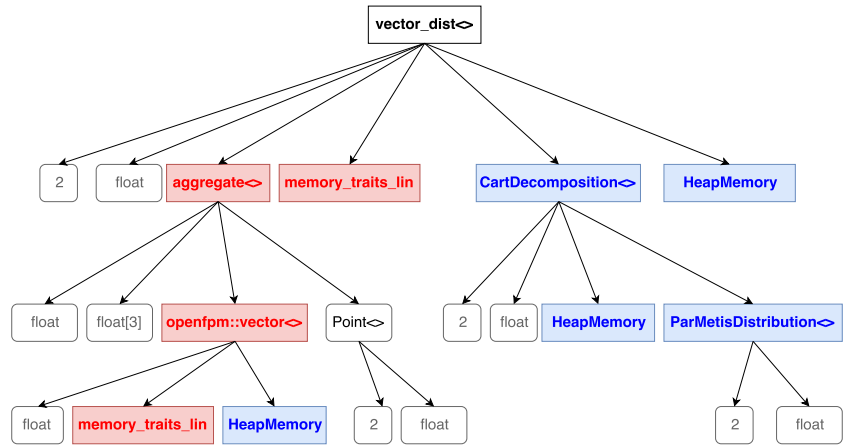

Figure 3: Tree expansion of the parametric data structure from Fig. 2b at compile-time. Introspection of the tree nodes allows the compiler to automatically generate implementations of all nodes and inline the code. Child nodes shown in blue are simply generated and inlined, whereas child nodes shown in red alter the implementation of their parents after introspection (non-local code injection).

minimum-surface sub-domains, this greedy heuristic is perfectly parallel and finds a sub-solution in linear time in the number of sub-sub-domains.

### 3.3. Parametric data structures

OpenFPM uses parametric data structures and templated C++ classes in order to adaptively generate optimized implementation code at compile-time. This is done for three reasons: First, it enables the internal implementation of data structures to adapt to the hardware platform for which they are being compiled, optimizing, e.g., the memory layout. Second, it avoids slow runtime polymorphism since dependencies can be resolved at compile-time and the corresponding code injected, leading to significant speed ups. Third, it renders the interface and semantics of the data structures independent of simulation parameters, such as the dimensionality of the computational domain or the floating-point precision used.

An example of such a parametric data structure is shown in Fig. 2 for a particle set. The parameters to the data structure are specified in the angle brackets. The example in Fig 2a defines a distributed vector (`vector_dist`) in 2D (first parameter is 2) with single-precision floating point numbers (the second parameter is "float"). The third parameter is the list (`aggregate`) of particle properties. In the example here, each particle stores a scalar float, an array of 3 floats, a double-precision `Point` object in a 3D space, and an OpenFPM Vector object (equivalent to `std::vector`) of floats. Parametric data structures can be arbitrarily nested and contain further parametric data structures within, as for example illustrated for the parametric `Point` object here.

There are additional, optional parameters that may be specified. In Fig 2b, some optional parameters of the `vector_dist` data structure are shown in addition to the mandatory parameters already present in Fig 2a. This includes parameters that define the memory layout of the data structure (here `memory_traits_lin` for a linear memory layout), the type of domain decomposition used to distribute this data structure (here a a decomposition into Cartesian blocks in 2D with float precision), the type of load balancer to be used on this data structure (here the distributed graph partitioning provided by the ParMetis library [22] in 2D), and the type of memory where the data inside this data structure are going to be allocated (here `HeapMemory`; this could also be GPU device memory, for example).

At compile-time, the template parameters of all data structures are expanded into a tree structure, as shown in Fig. 3 for the example from Fig. 2b. Each node in this tree contains the rules that specify how to generate the internal source code and call the child nodes. There are two types of child nodes: those that simply are called by the parent implementation (shown in blue in Fig. 3), and those that may alter the implementation of their parent node after introspection (shown in red in Fig. 3). The blue child nodes are inlined, such that all compiler optimizations apply. A parent node that has one or more red child nodes can alter its own implementation after introspection of the child nodes. This could, e.g., involve analyzing whenever a child node implements a certain method or defines specific attributes that can be used to transform the parent code to automatically adjust to the child node's interface, providing the possibility of non-local code injection. The compiler processes the tree recursively, until the internal implementations of all nodes have been generated.

As an example, consider data stored in an array of aggregates. The parent node `vector_dist` analyzes at compile-time whether any data type in the aggregate is a native object, like `int`, `double`, `float`, `float[3]`, etc. If this is not the case, the parent node checks at compile-time whether any C++ class contained in the aggregate defines a function `pack` to serialize its own data or a function `hasPointers` to indicate that the class points to data in another object. Depending on the outcome of these checks, different code for the parent node is generated. If at least one of the classes in the aggregate has `pack`, the parent node generates code to call this pack function for the specific variables contained in the aggregate. If none of the aggregate classes defines a pack function, and no pointers are present, the code is optimized to perform a single memory copy of the full array of aggregates. In case there are pointers, but no `pack` function is defined, code to display an error message at run-time is generated. If all of the classes in the aggregate define neither a `pack` function nor `hasPointers`, simple memory-copy code is generated, but a warning message is displayed to make the user aware that is not guaranteed to be safe.

Compile-time code injection, analysis, and introspection rely on the template engine of the C++ compiler. TMP is used for conditional code injection/inlining. While injected/inlined code is implementation-dependent, the meta-code required for analysis is independent of the specific data structure implementation and is encapsulated into the modular and re-usable parametric structures of OpenFPM.

Parametric data structures are also used by OpenFPM to manage memory and to control memory allocation patterns. Parameters are, e.g., available to control whether code is generated to allocate the contents of a data structure on the heap (see Fig 2b), on device memory such as on GPUs, or in pre-allocated external memory (e.g., transparent memory hoards). The code for the corresponding methods to allocate and delete instances of the data structure is automatically injected at compile-time. The same is also used to control memory layout (e.g., `memory_traits_lin` in Fig 2b) or to force a data structure to have a certain implementation (i.e., array of structs vs. struct of arrays).

In conclusion, the internal implementations of all OpenFPM data structures are automatically and adaptively generated at compile-time through a multi-level and modular source-code generation process relying on TMP. This opens the possibility to aggressively optimize the code, increase its flexibility and versatility, and to help the compiler exploit compile-time information to produce more scalable code. In addition, it also renders the code more compact, significantly reducing development and maintenance overheads.

*3.4. Mappings*

Mappings are OpenFPM communication abstractions [1]. They are provided in order to transparently distribute data according to a previously determined domain decomposition and to manage ghost layers. Each OpenFPM data structure can have its own domain decomposition and provides a method `map()` to distribute data according to this decomposition. The map method is typically used when particles have migrated across processor boundaries (local map), after distributed reading of data from a file (global map), or after the load balancer has altered the domain decomposition (global map). In a *local mapping*, processors only communicate to their neighbors. In this case, the map function internally sets up an asynchronous communication schedule, which is then executed using non-blocking point-to-point communications. This effectively avoids the need for an explicit communication schedule, e.g., using graph coloring, as done in synchronous environments [39]. The *global mapping* uses dynamic sparse data exchange (DSDE), as implemented by the non-blocking consensus (NBX) protocol [55].

Each OpenFPM data structure also provides methods to manage ghost layers, called *ghost mappings*. OpenFPM provides two types of ghost mappings: `ghost_get` and `ghost_put`. Ghost layers are populated with copies of the particles and/or mesh nodes from the overlapping regions of the neighboring processors using the `ghost_get` mapping. Calling `ghost_get` automatically populates the ghost layers of all processors without the user needing to know where the data came from, or how it has been communicated. This again uses an optimized internal communication schedule, as described above. The second ghost-related mapping is `ghost_put`. This mapping is used to send data from the ghost layer back to the source processor that stores the corresponding particle or mesh node. This is for example required when evaluating symmetric particle interactions that also alter the values of a ghost. In this case, the ghost contributions need to be sent back to the source processor where they are added to the local interaction results. The `ghost_put` mapping provides three different ways of merging the sent-back ghost contributions with the results on the source processor: sum the values on the source processor, replace the value on the source processor with the maximum of all incoming contributions, or merge all contributions into a list. The first is typically used to send symmetric interaction results back for aggregation. The second could be used for collision detection, while the third is typically the case when managing contact lists in discrete-particle simulations. In addition to these three pre-defined `ghost_put` merging operations, the user may also implement own operators as C++ template functors.

All mappings are internally implemented using non-blocking MPI communication. In all mappings, it is also possible to only map a subset of the properties of the affected particles or mesh nodes. This is done by specifying the list of properties that are to be mapped as an optional template parameter to the mapping methods.

*3.5. Dynamic load balancing*

Domain decomposition leads to an initially load-balanced and communication-minimized distribution of work and data across processors. During the course of a simulation, however, particle movement, dynamic resolution refinement, and particle evolution may lead to progressively worse load balance and increasing communication overhead. OpenFPM therefore provides runtime load re-balancing. Again, like in the initial domain decomposition, the task is formulated as a graph-partitioning problem that is approximately solved using ParMetis [22]. Load re-balancing is done on the level of sub-sub-domains and is followed by another sub-domain creation step, but avoids *de novo* domain decomposition. In addition, the current decomposition graph is used as a soft constraint for the new graph decomposition in order to minimize load migration. An additional migration cost $m_i$ is assigned to each sub-sub-domain, corresponding to the amount of data that needs to be communicated if that sub-sub-domain changes processor. This cost term is included into the graph-decomposition problem in order to obtain re-balanced processor assignments that are close to the previous one and that trade off the migration cost of transferring sub-sub-domains between processors against the gain in load balance. The data-transfer cost is linearly discounted over the number of simulation time steps since the last re-balancing. The time-points when re-balancing is beneficial, i.e., when the cost of re-balancing is amortized by the gain in load balance, are automatically determined using the Stop-At-Rise (SAR) heuristic [56] or can be specified by the user program.

*3.6. Iterators*

Because of the non-trivial distribution of data across processors, looping over particles or mesh nodes in a transparent and cache-friendly manner requires additional information. In order to hide this complexity from the user, OpenFPM provides transparent iterators for particles and meshes.

On a mesh, the iterators are multi-dimensional (i.e., multi indices) and follow the natural topology of the mesh. On particles, iterators are one-dimensional. Since the indexing order of the particles is inconsequential for the simulation result (see Section 2), the user can select iterators that automatically create cache-friendly loops, which may re-sort or re-order the particles in a more memory-efficient way. Separate iterators are provided for interior particles and interior mesh nodes and for ghost particles and ghost mesh nodes, which allows overlapping communication with computation by iterating over interior elements while the ghosts are mapped in the background, before iterating also over the ghosts. Examples of iterators for both particles and meshes are shown in the code examples in Section 4 below.

*3.7. File I/O, check-point restart*

The distributed data structures provided by OpenFPM can be saved to file using the portable binary HDF5 file format [57]. This is done by calling the OpenFPM method `save()`, irrespective of how the data are distributed across processors. Files written in this way can be read back using the OpenFPM method `load()`. Since HDF5 is a parallel file format, reading can be done on any number of processors, independent from the number of processors used to save the file, and the domain decompositions at the time of saving and loading may be different. OpenFPM HDF5 files can therefore also directly be used to restart a simulation from a previously saved state (check-point restart).

The HDF5 format of OpenFPM internally serializes all data structures, which allows arbitrary C++ objects and nested parametric data structures to be transparently supported. For this, OpenFPM provides an encapsulated serialization/de-serialization sub-system that is used by each processor locally in order to serialize the local pieces of a distributed data structure into a so-called "chunk". The chunks from all processors are then saved contiguously into a parallel HDF5 file. Meta-data are automatically added by the serializer to allow de-serialization on other numbers of processors and other domain decompositions. For this, chunks are read in parallel by individual processors and, after distributed de-serialization, mapped onto the new domain decomposition. This map-after-read strategy is preferred, as it causes data to be read/written in large contiguous blocks rather than in multiple smaller random reads, producing less load in the I/O subsystem.

In addition to HDF5 output, distributed data can also be saved to files in VTK format [58] using the OpenFPM method `write()`. This enables direct visualization of the particle and mesh data, e.g., in Paravew [59], an open-source scientific data visualization software that natively reads VTK files. While OpenFPM HDF5 files cannot directly be visualized in Paraview, they can easily be converted to VTK files using `load()` to reload the HDF5 filed into distributed data-structures followed by `write()` to export to VTK. VTK output and Paraview were used to generate all visualizations in Section 4 of this paper.

## 4. Results

We demonstrate the use, performance, and scalability of OpenFPM in a number of test cases from different application domains. We compare with a respective state-of-the-art code in each application domain, demonstrating that a generic framework like OpenFPM can reach and in some cases even exceed the performance of application-specific codes that matured over many years. All tests are performed on the computer cluster of the Center for Information Services and High-Performance Computing of TU Dresden. Each node of the cluster is equipped with two 2.5 GHz 12-core Intel Xeon E5-2680v3 CPUs (total 24 cores per node) sharing 64 GB of RAM. The cluster interconnect is an Infiniband network with 40 Gb/s bandwidth. The machine is running RedHat Enterprise Linux (RHEL) Server release 6.9 (Santiago) as operating system. For all tests, OpenFPM was compiled using GCC g++ version 7.1.0 and linked against the OpenMPI 3.0.0 [13] implementation of the MPI standard version 3 [12].

On the benchmark machine, each processor (socket) has its own, independent memory bus. The cores within each processor, however, share the memory bandwidth. The results of a concurrent memory-read benchmark are shown in Table 1. The memory bandwidth reduces from 11.5 GB/s when using only 1 core to 5.0 GB/s when using all 12 cores of a processor in parallel. Running on 16 cores with rank-affinity binding will, e.g., result in the first 12 cores running at 5.0GB/s and the remaining 4 cores at 10 GB/s. We use rank-affinity binding in all benchmarks below. Perfect scalability is thus not to be expected on this hardware.

### 4.1. Molecular dynamics

We first consider a classic Molecular Dynamics (MD) application simulating a Lennard-Jones fluid. In this simulation, particles represent atoms that interact according to the pairwise

| #cores/processor: | 1 | 2 | 4 | 8 | 12 |
|---|---|---|---|---|---|
| bandwidth GB/s: | 11.5 | 10.8 | 10.0 | 8.3 | 5.0 |

Table 1: Memory bandwidth on the benchmark machine when using different numbers of cores on the same processor.

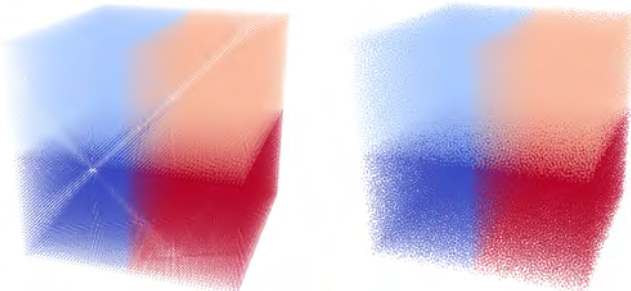

Figure 4: Particle configurations at the start of the simulation (left) and after 5000 time steps (right) for the Lennard-Jones molecular dynamics test case. The system was thermally equilibrated after 1000 time steps. In this example, a decomposition into four sub-domains is used, indicated by different colors. Each particle is plotted as a dot with the color of the respective sub-domain.

Lennard-Jones potential:

$$V_{\text{LJ}}(r) = 4\epsilon \left[ \left(\frac{\sigma}{r}\right)^{12} - \left(\frac{\sigma}{r}\right)^{6} \right]$$

as a function of the distance $r$ between the two interacting particles. The parameters $\sigma$ and $\epsilon$ define the zero-crossing and the well depth of the potential, respectively. We implement the simulation in OpenFPM using OpenFPM's implementation of Verlet lists [46] for the particles to efficiently find their interaction partners. We exploit symmetry in the interactions, i.e., we compute every interaction pair only once. This also requires changing the values of ghost particles, for which we use OpenFPM's ghost_put mapping (see Section 3.4).

We compare the OpenFPM-based implementation with LAMMPS [28], a well-established and highly optimized parallel MD code. We start the simulation with 216,000 particles initialized on a regular Cartesian $60^3$ mesh. This initial particle configuration is shown in the left panel of Fig. 4 for a Cartesian domain decomposition with four sub-domains indicated by different colors. We then first equilibrate the system and then simulate its time dynamics using the symplectic Velocity-verlet time-stepping scheme [46] with step size $\delta t = 0.01$. For the Lennard-Jones potential, we use $\sigma = 0.1$ and $\epsilon = 1$. Periodic boundary conditions are imposed in all three coordinate directions. We run the simulation for 5000 time steps. The final configuration of particles is shown in the right panel of Fig. 4. The time courses of the kinetic, potential, and total energies of the system as computed by LAMMPS and OpenFPM were identical (data not shown) and the total energy was conserved, hence validating the simulation. Equilibration was finished after about 1000 time steps.

| #cores | OpenFPM (seconds) | LAMMPS (seconds) | OpenFPM (Efficiency) | LAMMPS (Efficiency) |
|---|---|---|---|---|
| 1 | 1010.69 ± 1.58 | 976.10 ± 3.30 | 100% | 100% |
| 4 | 262.55 ± 0.80 | 257.00 ± 6.48 | 96.2% | 94.9% |
| 8 | 143.81 ± 0.26 | 137.10 ± 0.31 | 87.8% | 89.0% |
| 16 | 77.10 ± 0.40 | 73.00 ± 0.46 | 81.9% | 83.6% |
| 24 | 52.70 ± 0.27 | 49.89 ± 0.17 | 79.9% | 81.5% |
| 48 | 29.70 ± 0.12 | 28.98 ± 0.22 | 70.9% | 70.2% |
| 96 | 15.16 ± 0.11 | 15.64 ± 0.60 | 69.4% | 65.0% |
| 192 | 8.07 ± 0.16 | 8.22 ± 0.32 | 65.2% | 61.8% |
| 384 | 4.73 ± 0.16 | 4.66 ± 0.17 | 55.6% | 54.5% |
| 768 | 3.15 ± 0.09 | 3.37 ± 1.10 | 41.8% | 37.7% |
| 1536 | 2.20 ± 0.24 | 1.93 ± 0.77 | 29.9% | 32.9% |

Table 2: Molecular dynamics benchmark results. We report wall-clock execution times (mean ± standard deviation over 10 independent runs) and parallel efficiencies of the OpenFPM client compared with LAMMPS [28] for a strong scaling from 1 to 1536 processor cores simulating 216,000 Lennard-Jones particles in the unit cube over 5000 time steps (see Fig. 4).

The OpenFPM-based simulation can be implemented in less than 40 lines of C++ code (not counting comments), as shown in Listing 4.1. In the example listing, lines 10–15 define the pairwise Lennard-Jones interaction between the particles. The main program starts by initializing OpenFPM in line 19 and defining the size of the simulation domain in line 28 (here: the unit cube), the boundary conditions in line 29 (here: periodic in all three directions), and the size of the ghost layers in line 30 (here: given by the interaction cut-off radius r_cut). The particle interaction object is instantiated in line 33 based on the definition of the interaction in lines 10–15. Line 36 allocates a distributed vector to store the particles. The particle position consists of 3 doubles (first two template parameters), and the particle properties are an aggregate containing two 3D points/vectors of doubles. Lines 6 and 7 define that the first property (i.e, property 0) is the velocity and the second (i.e, property 1) is the force. The particles are then initialized on a regular Cartesian lattice in line 37, where the number of grid points (sz) has been defined in line 27 (here: 60 in each direction). Lines 41–43 define convenient aliases for the particle position, velocity, and force that can be used in expressions like the one in line 58 to simplify notation. Line 50 creates the symmetric particle cell lists for fast neighbor access and uses them in line 51 to compute the initial forces using symmetric particle interactions (omitting the _sym specifier would compute the interactions without exploiting symmetry. The _in specifier tells OpenFPM to only compute interactions for internal (i.e., non-ghost) particles. Lines 54–73 contain the simulation time loop using the two-step velocity Verlet symplectic time integrator (lines 58, 59, 72). Lines 63 and 64 contain communication operations. Line 63 performs a mapping to migrate particles that have moved across processor boundaries, and line 64 performs a ghost-get mapping for only the particle positions (empty properties list <>). Line 76 finalizes the OpenFPM Library at the end of the program.

Table 2 compares the performance of the OpenFPM-based implementation using Verlet lists with LAMMPS for a strong scaling, i.e., distributing the fixed number of 216,000 particles across an increasing number of processors. The absolute wall-clock time per time step is below 1 second even on a single core. On 1536 cores, a simulation time step is completed in 0.5 ms. Despite the fact that OpenFPM is a general-purpose particle-mesh library and is not limited to MD, its performance is almost as good as that of the highly optimized LAMMPS.

[Listing 4.1: C++ code for Lennard-Jones molecular dynamics using OpenFPM]

```cpp
///// define parameters
double sigma12, sigma6, epsilon = 1.0, sigma = 0.1;   // parameters of the potential
double dt = 0.0005, r_cut = 3.0*sigma;                // parameters of the simulation
double r_cut2;

constexpr int velocity_prop = 0;  // velocity is the first particle property
constexpr int force_prop = 1;     // force is the second particle property

///// Define Lennard-Jones interaction to be used in applyKernel_in_sym
DEFINE_INTERACTION_3D(ln_force)
    Point<3,double> r = xp - xq;
    double rn = norm2(r);
    if (rn > r_cut2) return 0.0;
    return 24.0*epsilon*(2.0*sigma12/(rn*rn*rn*rn*rn*rn*rn)-sigma6/(rn*rn*rn*rn))*r;
END_INTERACTION

int main(int argc, char* argv[]) {
    ///// Initialize OpenFPM
    openfpm_init(&argc,&argv);

    ///// Initialize constants
    sigma6 = pow(sigma,6), sigma12 = pow(sigma,12);
    r_cut2 = r_cut*r_cut;

    ///// Define initialization grid, simulation box, periodicity
    ///// and ghost layer
    size_t sz[3] = {60,60,60};
    Box<3,float> box({0.0,0.0,0.0},{1.0,1.0,1.0});
    size_t bc[3]={PERIODIC,PERIODIC,PERIODIC};
    Ghost<3,float> ghost(r_cut);

    ///// Lennard-Jones potential object used in applyKernel_in
    ln_force lennard_jones;

    ////// Define particles and initialize them on a grid
    vector_dist<3,double,aggregate<Point<3,double>,Point<3,double>>> particles(0,box,bc,
        ghost);
    Init_grid(sz,particles);

    ///// Define aliases for the particle force, velocity, and position
    ///// to simplify notation
    auto force = getV<force_prop>(particles);
    auto velocity = getV<velocity_prop>(particles);
    auto position = getV<PROP_POS>(particles);

    ///// initialize all particle velocities to zero
    velocity = 0;

    ///// Generate the cell lists and compute the initial forces using the Lennard-Jones
    ///// potential evaluated with exploiting symmetry
    auto NN = particles.getCellListSym(r_cut);
    force = applyKernel_in_sym(particles,NN,lennard_jones);

    ///// Time loop
    for (size_t i = 0; i < 10000 ; i++) {
        ///// 1st step of velocity Verlet time integration
        ///// v(t + 1/2*dt) = v(t) + 1/2*force(t)*dt
        ///// x(t + dt) = x(t) + v(t + 1/2*dt)
        velocity = velocity + 0.5*dt*force;
        position = position + velocity*dt;

        ///// communicate particles that have crossed processor boundaries and
        ///// update the ghost layers for all properties (empty props list)
        particles.map();
```

```
64            particles.ghost_get<>();
65
66            // Calculate the forces at t + dt
67            particles.updateCellListSym(NN);
68            force = applyKernel_in_sym(particles,NN,lennard_jones);
69
70            ///// 2nd step of velocity Verlet time integration
71            ///// v(t+dt) = v(t + 1/2*dt) + 1/2*force(t+dt)*dt
72            velocity = velocity + 0.5*dt*force;
73        }
74
75        ///// Finalize OpenFPM and deallocate all memory
76        openfpm_finalize();
77    }
```

### 4.2. Smoothed-particle hydrodynamics

Smoothed-Particle Hydrodynamics (SPH) is a widely used method for simulating continuous models of fluid dynamics. Due to its simplicity and flexibility in modeling complex fluid properties and free fluid surfaces, it is preferentially used to model multi-phase flows and fluid-structure interaction [60, 61].

We use OpenFPM to implement a weakly compressible SPH solver for the Navier-Stokes equations, where each particle $p$ has a velocity $v_p$, a pressure $P_p$, and a density $\rho_p$. The evolution of these particle properties is governed by [34]:

$$\frac{dv_p}{dt} = -\sum_{q \in \mathcal{N}(p)} m_q \left( \frac{P_p + P_q}{\rho_p \rho_q} + \Pi_{pq} \right) \nabla W(x_q - x_p) + g \tag{1}$$

$$\frac{d\rho_p}{dt} = \sum_{q \in \mathcal{N}(p)} m_q v_{pq} \cdot \nabla W(x_q - x_p) \tag{2}$$

$$P_p = b \left[ \left( \frac{\rho_p}{\rho_0} \right)^\gamma - 1 \right] \tag{3}$$

$$b = \frac{1}{\gamma} c_{\text{sound}}^2 |g| h_{\text{swl}} \rho_0, \tag{4}$$

where $h_{\text{swl}}$ is the maximum height of the fluid, $\gamma = 7$, and $c_{\text{sound}} = 20$ [34]. Here, $\mathcal{N}(p)$ is the set of all particles within a cutoff radius of $2\sqrt{3}h$ from $p$, where $h$ is the distance between nearest neighbors. $W(x)$ is the classic cubic SPH kernel [34] and $g$ is the gravitational acceleration. The relative velocity between particles $p$ and $q$ is $v_{pq} = v_p - v_q$, $\nabla W(x_q - x_p)$ is the analytical gradient of the kernel $W$ centered at particle $p$ and evaluated at the location of particle $q$. Equation 3 is the equation of state that links the pressure $P_p$ with the density $\rho_p$, where $\rho_0$ is the density of the fluid at $P = 0$. $\Pi_{pq}$ is the viscosity term defined as:

$$\Pi_{pq} = \begin{cases} -\frac{\alpha c_{pq}^- \mu_{pq}}{\rho_{pq}^-} & v_{pq} \cdot r_{pq} > 0 \\ 0 & v_{pq} \cdot r_{pq} < 0 \end{cases} \tag{5}$$

with constants defined as: $\mu_{pq} = \frac{h v_{pq} \cdot r_{pq}}{r_{pq}^2 + \eta^2}$ and $c_{pq}^- = \sqrt{g \cdot h_{swl}}$.

We use the OpenFPM-based implementation to simulate a water column impacting onto a fixed obstacle. This "dam break" scenario is a standard test case for SPH simulation codes. A visualization of the OpenFPM result at three different time points is shown in Fig. 5. We compare the results and performance with those obtained using the popular open-source SPH code DualSPHysics [26]. The publicly available version of DualSPHysics only supports shared-memory multi-core platforms and GPGPUs, which is why we limit comparisons to these cases.

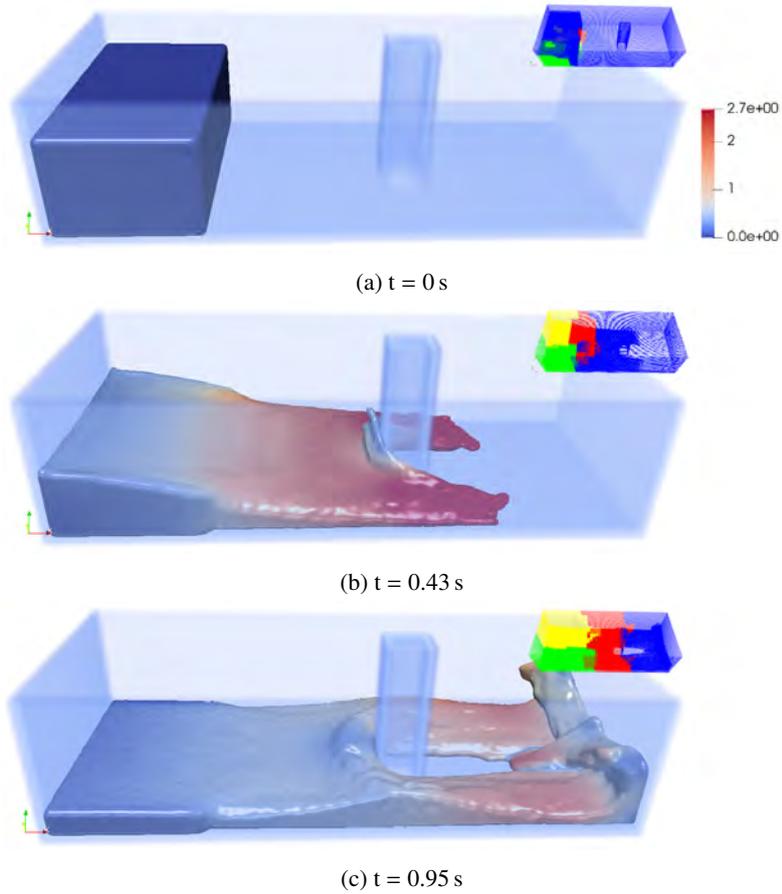

(a) t = 0 s

(b) t = 0.43 s

(c) t = 0.95 s

Figure 5: Visualization of the SPH dam-break simulation. We show the fluid particles at times 0, 0.43, and 0.95 s of simulated time, starting from a column of fluid in the left corner of the domain as shown. We use the OpenFPM SPH to solve the weakly compressible Navier-Stokes equations with the equation of state for pressure as given in Eqs. 1–3. The figure shows a density iso-surface indicating the fluid surface with color indicating the fluid velocity magnitude. The small insets show the distribution of the domain onto 4 processors with different processors shown by different colors. The dynamic load balancing of OpenFPM automatically adjusts the domain decomposition to the evolution of the simulation in order to maintain scalability.

The present SPH implementation based on OpenFPM uses the same algorithms as DualSPHysics [26], with identical initialization, boundary conditions, treatment of the viscosity term, and Verlet time-stepping [46] with dynamic step size. The results are therefore directly comparable. In this test case, particles are not homogeneously distributed across the domain, and they significantly move during the simulation. Therefore, this provides a good showcase for the dynamic load balancing capability of OpenFPM.

We validate our simulation by calculating and comparing the velocity and pressure profiles at multiple points between OpenFPM and DualSPHysics [26]. We find that all pressure and velocity profiles are identical (not shown). We measure the performance of the OpenFPM-based implementation in comparison with the DualSPHysics code running on 24 cores of a single cluster node. We simulate the dam-break case with 171,496 particles until a physical time of 1.5 seconds. The OpenFPM code completes the entire simulation in about 500 seconds, whereas DualSPHysics requires about 950 seconds. The roughly two-fold better performance of OpenFPM is possibly attributed to the use of symmetry when evaluating the interactions and the use of optimized Verlet lists, which do not seem to be exploited in DualSPHysics [26].

Since DualSPHysics is mainly optimized for use on GPGPUs, we also compare the OpenFPM-based implementation in distributed-memory mode with DualSPHysics running on a GPGPU. The benchmark is done with 15 million SPH particles using a nVidia GeForce GTX1080 GPU. The OpenFPM code reached the same performance when running on around 270 CPU cores of the benchmark machine and was faster when using more cores. This shows that OpenFPM can reach GPU performance on moderate numbers of CPU cores without requiring specialized CUDA code.

We also use this test case to profile OpenFPM with respect to the fraction of time spent computing, communicating, and load-balancing. The results are shown in Table 3 for different numbers of particles on 1536 processors, hence testing the scalability of the code to large numbers of particles. The small insets in Fig. 5 show how the domain decomposition of OpenFPM dynamically adapts to the evolving particle distribution by dynamic load re-balancing (see Section 3.5). In this example, the load distribution strongly changes due to the large bulk motion of the particles. The dynamic load-balancing routines of OpenFPM consume anywhere between 5 and 25% of the total execution time, but their absolute runtime is independent of the number of particles. Therefore, the relative fraction of communication and load-balancing decreases for increasing number of particles, whereas the average imbalance remains roughly constant due to the dynamic load balancing. Since load balancing and communication are not required when running on a single core, the percentage of time spent computing (second column in Table 3) can directly be interpreted as the parallel efficiency of the code on 1536 cores, which is, as expected, increasing with problem size.

### 4.3. Finite-difference reaction-diffusion code

As a third showcase, we consider a purely mesh-based application, namely a finite-difference code to numerically solve a reaction-diffusion system. Reaction-diffusion systems are widely studied due to their ability to form steady-state concentration patterns, including Turing patterns [62]. A particularly known example is the Gray-Scott system [63, 64, 65, 66], which produces a rich variety of patterns in different parameter regimes. It is described by the following

| #particles | Computation (%) | Imbalance (%) | DLB (%) | Communication (%) | Time (s) |
|---|---|---|---|---|---|
| 0.46M | 19.2% | 15.1% | 25.7% | 40.0% | 148.40 |
| 1.20M | 32.3% | 30.4% | 10.5% | 26.8% | 346.83 |
| 4.00M | 50.0% | 22.4% | 10.6% | 17.0% | 1424.64 |
| 10.78M | 64.0% | 21.2% | 6.81 % | 8.09% | 4666.78 |
| 14.63M | 65.5% | 21.5% | 5.38% | 7.71% | 7035.01 |

Table 3: Percentage of the total runtime spent on different tasks by OpenFPM for the SPH dam-break simulation on 1536 cores using different numbers of particles (1st column). The computation time is the average wall-clock time across processors spent on local computations, while the load imbalance is given by the difference between the maximum wall-clock time across processors and the average. Communication is the time taken by all mappings together, and DLB (dynamic load balancing) is the time taken to decompose the problem and assign sub-domains to processors. The last column gives the total runtime of the simulation until a simulated time of 1.5 s.

set of partial differential equations:

$$\begin{aligned}\frac{\partial u}{\partial t} &= D_u \nabla u - uv^2 + F(1-u) \\ \frac{\partial v}{\partial t} &= D_v \nabla v + uv^2 - (F+k)v\,,\end{aligned} \quad (6)$$

where $D_u$ and $D_v$ are the diffusion constants of the two species with concentrations $u$ and $v$, respectively. The parameters $F$ and $k$ determine the type of pattern that is formed.

We implement an OpenFPM-based numerical solver for these equations using second-order centered finite-differences on a regular Cartesian mesh in 3D of size $256^3$. We compare the performance of the OpenFPM-based implementation with that of an efficient AMReX-based solver [30]. Even though AMReX is a multi-resolution adaptive mesh-refinement code, we still use it as a benchmark also in the present uniform-resolution case because it is highly optimized. However, AMReX requires the user to tune the maximum grid size for data distribution [30]. If we choose the maximum grid size too large, AMReX does not have enough granularity to parallelize. If it is chosen too small, scalability is impaired by a larger ghost-layer communication overhead. This parameter for AMReX was determined manually in order to ensure that the number of sub-grids is always larger than the number of processor cores used. The actual values used are given in the last column of Table 4. OpenFPM does not require the user to set such a parameter, as the domain decomposition is determined automatically. For both AMReX and OpenFPM, we use MPI-only parallelism in order to compare the results.

For the benchmark simulations, we use the following parameter values: $D_u = 2 \cdot 10^{-5}, D_v = 10^{-5}$, varying $k$ and $F$ as given in the legends of Fig. 6 to produce different patterns. To validate the simulation, we reproduce the nine patterns classified by Pearson [67], with visualizations shown in Fig. 6.

An OpenFPM source-code example of applying a simple 5-point finite-difference stencil to a regular Cartesian mesh is shown in Listing 4.3. The stencil is defined in line 2 as an OpenFPM grid key array with relative grid coordinates. Here, the stencil object is called `star_stencil_2D` and consists of 5 points. In line 5, a mesh iterator is created for this stencil as applied to the mesh object `Old`. Lines 7–24 then loop over all mesh nodes and apply the stencil. The expression of the stencil (lines 18–20) issimplified by first defining aliases for the shifted nodes in lines 10–14, albeit this is not necessary.

[Listing: 4.3: OpenFPM code example for stencil operations on a regular Cartesian mesh]

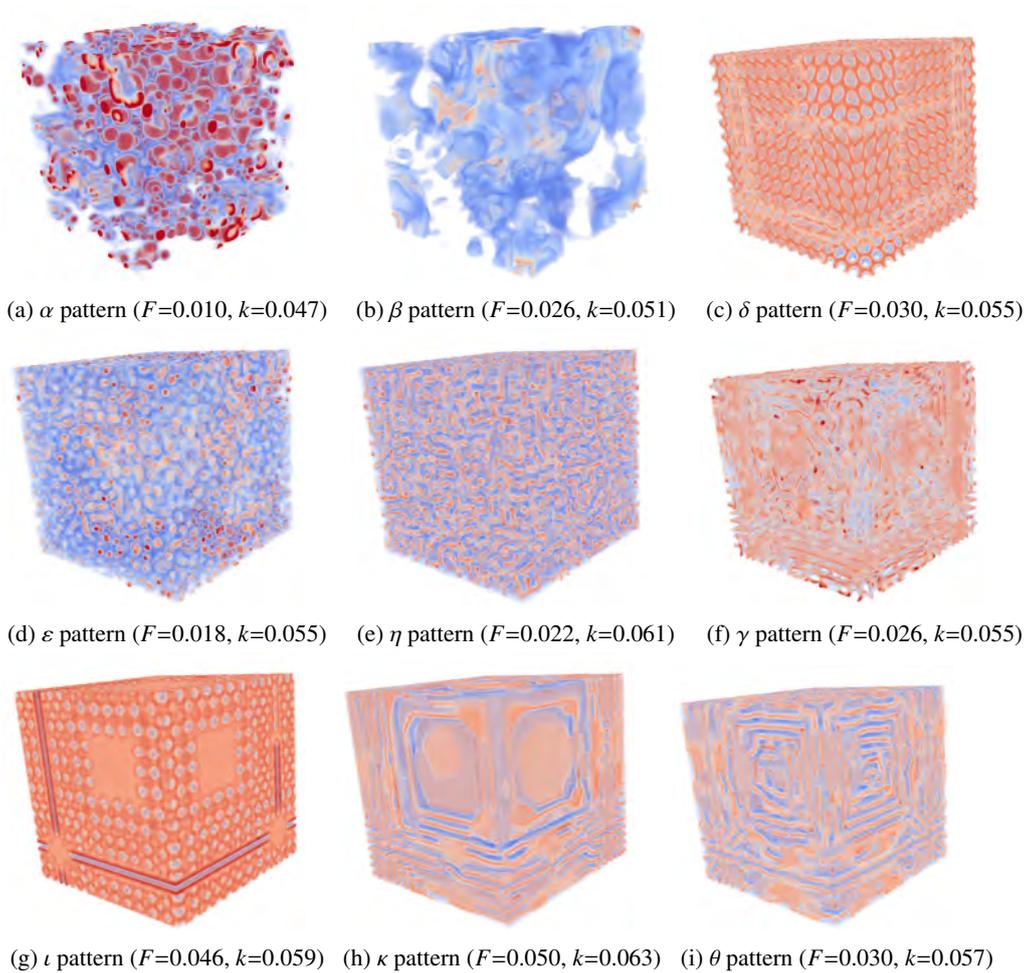

Figure 6: Visualizations of the OpenFPM-simulations of nine steady-state patterns produced by the Gray-Scott reaction-system in 3D [67] for different values of the parameters $F$ and $k$.

```
1   ///// finite-difference stencil definition
2   static grid_key_dx<2> star_stencil_2D[5] = {{0,0},{-1,0},{+1,0},{0,-1},{0,+1}};
3
4   ///// create an iterator for the stencil on the mesh "Old"
5   auto it = Old.getDomainIteratorStencil(star_stencil_2D);
6
7   while (it.isNext()) {
8       ///// define aliases for center, minus-x, plus-x, minus-y, plus-y.
9       ///// The template parameter is the stencil element.
10      auto Cp = it.getStencilGrid<0>();
11      auto mx = it.getStencilGrid<1>();
12      auto px = it.getStencilGrid<2>();
13      auto my = it.getStencilGrid<3>();
14      auto py = it.getStencilGrid<4>();
15
16      ///// apply the stencil to field U on mesh "Old" and store
```

```
17        ///// the result in the field U on mesh "New"
18        New.get<U>(Cp) = Old.get<U>(Cp) +
19            (Old.get<U>(my)+Old.get<U>(py)+Old.get<U>(mx)+Old.get<U>(px) -
20            4.0*Old.get<U>(Cp));
21
22        ///// Move to the next mesh node
23        ++it;
24  }
```

The performance of OpenFPM compared to AMReX is shown in Table 4 and Fig. 7. OpenFPM scales slightly better than AMReX, with wall-clock times in the same range. Both codes saturate at the same wall-clock time for large numbers of cores (Fig. 7). Both AMReX and OpenFPM use mixed C++/Fortran code for this benchmark, with all stencil iterations implemented in Fortran. Because Fortran provides native support for multi-dimensional arrays, it produces more efficient assembly code than C++. In our tests, a fully C++ version was about 20% slower than the hybrid C++/Fortran. We note that the present benchmark problem is relatively small ($256^3$ mesh nodes), which is why strong scaling saturates already at about 24 cores.

| #cores | OpenFPM (seconds) | AMReX (seconds) | AMReX param |
|---|---|---|---|
| 1 | 393.1 ± 1.3 | 388.5 ± 1.5 | 256 |
| 2 | 207.5 ± 1.3 | 265.0 ± 0.8 | 128 |
| 4 | 105.8 ± 1.3 | 144.8 ± 0.3 | 128 |
| 8 | 65.1 ± 2.1 | 106.6 ± 2.6 | 128 |
| 12 | 65.6 ± 2.6 | 90.9 ± 5.0 | 64 |
| 16 | 57.6 ± 1.9 | 173.6 ± 3.6 | 64 |
| 20 | 56.8 ± 2.0 | 66.0 ± 1.7 | 64 |
| 24 | 60.5 ± 0.3 | 60.9 ± 4.0 | 64 |

Table 4: Performance of the OpenFPM finite-difference code compared with AMReX [30]. Times are given in seconds as mean±standard deviation over 10 independent runs for a fixed problem size of $256^3$ mesh nodes (strong scaling). The grid-size parameters used for AMReX are given in the last column.

*4.4. Vortex Methods*

As a fourth showcase, and in order to show how OpenFPM handles hybrid particle-mesh problems, we consider a full vortex-in-cell [49] code, a hybrid particle-mesh method to numerically solve the incompressible Navier-Stokes equations in vorticity form with periodic boundary conditions. These equations are:

$$\frac{D\boldsymbol{\omega}}{Dt} = (\boldsymbol{\omega} \cdot \nabla)\boldsymbol{u} + \nu \Delta \boldsymbol{\omega}$$
$$\Delta \boldsymbol{\psi} = \nabla \times \boldsymbol{u} = \boldsymbol{\omega}, \quad (7)$$

with $\boldsymbol{\omega}$ the vorticity, $\boldsymbol{\psi}$ the vector stream function, $\nu$ the viscosity, and $\boldsymbol{u}$ the velocity of the fluid. The operator $\frac{D}{Dt}$ denotes a Lagrangian (material) time derivative [49]. We numerically solve these equations using an OpenFPM-based implementation of the classic vortex-in-cell method as given in Algorithm 1 with two-stage Runge-Kutta time stepping. Particle-mesh and mesh-particle interpolations use the moment-conserving $M'_4$ interpolation kernel [34].

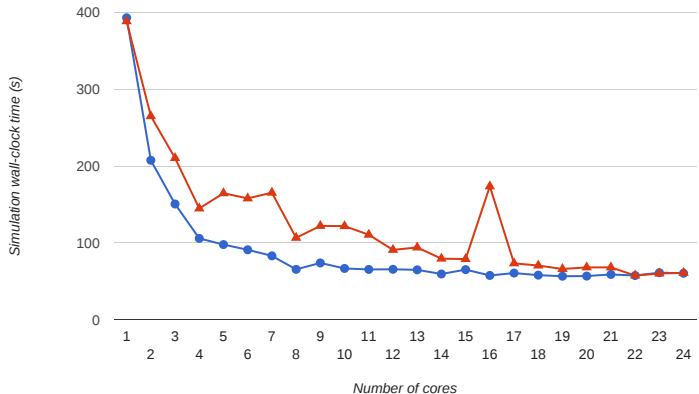

Figure 7: Scalability of the OpenFPM finite-difference code (blue circles) in comparison with AMReX [30] (red triangles) for a strong scaling. Shown is the wall-clock time in seconds to complete 5000 time steps of the Gray-Scott finite-difference code (5-point stencil) on a $256^3$ uniform Cartesian grid using different numbers of cores.

We run a simulation that reproduces previous results of a self-propelling vortex ring [68]. The vortex ring is initialized on a grid of size $1600 \times 400 \times 400$ using

$$\omega_0 = \frac{\Gamma}{\pi\sigma^2} e^{-s/\sigma}, \tag{8}$$

where $s^2 = (z - z_c)^2 + [(x - x_c)^2 + (y - y_c)^2 - R]$, with $R = 1$, $\sigma = R/3.531$, and the domain $(0 \ldots 5.57, 0 \ldots 5.57, 0 \ldots 22.0)$. We set $\Gamma = 1$, and $x_c = 2.785$, $y_c = 2.785$, $z_c = 2.785$ as the center of the torus defining the initial vortex ring.

A Runge-Kutta time-stepping scheme of order 2 is used with fixed step size $\delta t = 0.0025$. All differential operators are discretized using second-order symmetric finite differences on the mesh. We use 256 million particles distributed across 3072 processors to simulate the behavior of the vortex ring at Reynolds number Re = 3750 until final time $t = 225.5$. VTK files are written by OpenFPM and directly visualized using Paraview [59]. We observe the same patterns and structures for the ring as in Ref. [68], see Fig. 8.

The performance and scalability of the OpenFPM code are limited by the linear system solver required for computing the velocity from the vorticity on the mesh, i.e., by solving the Poisson equation. In this benchmark, OpenFPM internally uses a solver provided by the PetSc library [51]. We benchmark the parallel scalability of the solver and the overall code in a weak scaling starting from a $109 \times 28 \times 28$ mesh on 1 processor up to $1207 \times 317 \times 317$ mesh nodes on 1536 processors. We separately time the efficiency of the PetSc solver and of the OpenFPM parts of the code (particle-mesh/mesh-particle interpolation, remeshing, time integration, right-hand side evaluation). The results are shown in Fig. 9. Within a cluster node (1...24 cores), the decay in efficiency can be explained by the shared memory bandwidth (see Table 1). PetSc shows another marked drop in efficiency when transitioning from one cluster node to two nodes (48 cores). After that, the efficiency remains stable until 768 cores, when it starts to slowly drop again.

To put these results into perspective, we compare the particle-mesh interpolation part of the code with the corresponding part of a PPM-based hybrid particle-mesh vortex code previously used [39]. We only compare this part of the code in order to exclude differences between PetSc and the own internal solvers of PPM. Interpolating two million particles to a $128^3$ mesh takes 0.41 s in OpenFPM and 3.4 s in PPM on a single core. Performing a weak scaling starting from a $128^3$ mesh on 1 processor, the OpenFPM particle-mesh interpolation reaches a parallel efficiency of 75% on 128 cores (16 nodes using 8 cores of each node). This is comparable with the scalability of PPM on the same test problem (see Fig. 13 of Ref. [39]).

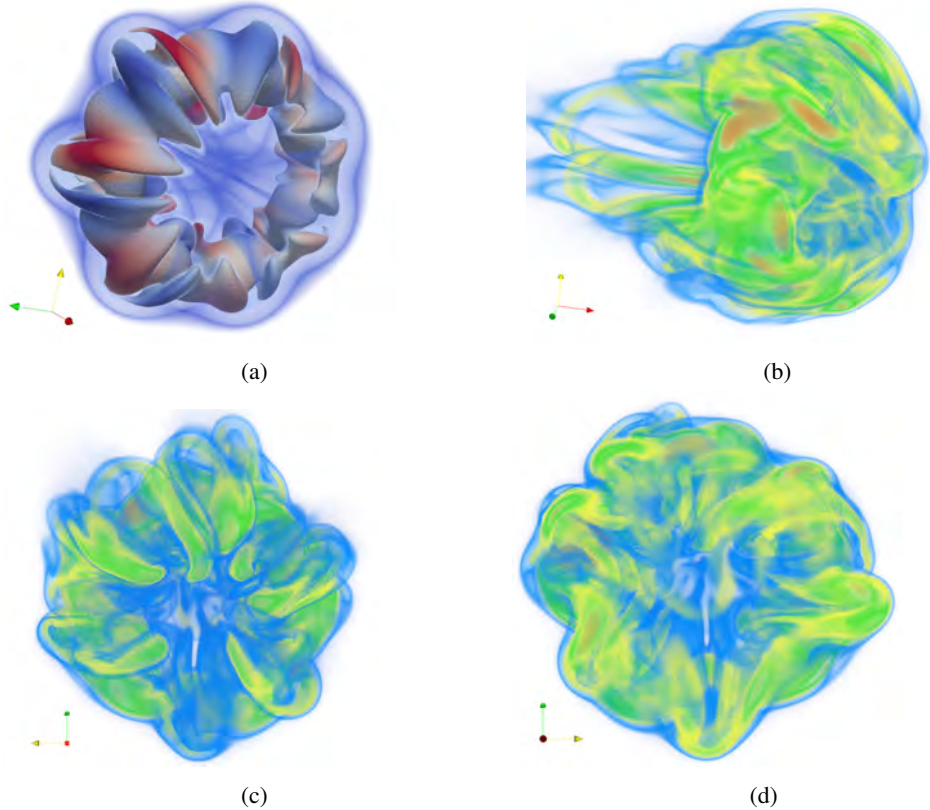

Figure 8: Visualization of the OpenFPM simulation of a vortex ring at Re=3750 using a hybrid particle-mesh Vortex Method (Algorithm 1) to solve the incompressible Navier-Stokes equations with 256 million particles on 3072 processors. Results are visualized for $t = 195.5$ when the ring is just about to become turbulent. (a) The iso-surfaces of vorticity highlight the tubular dipole structures in the vortex ring. Color corresponds to the $x$-component of the vorticity with red indicating positive signs and blue negative signs. (b)–(d) Three different views of a volume rendering of four vorticity bands: orange is $\|\omega\|^2 = 3.239\ldots2.3$, green is $\|\omega\|^2 = 1.16\ldots1.372$, yellow is $\|\omega\|^2 = 0.7\ldots0.815$, and blue is $\|\omega\|^2 = 0.3\ldots0.413$.

*4.5. Discrete element methods*

Discrete element methods (DEM) are important for the study of granular materials, in particular for determining effective macroscopic dynamics for which the governing equations are not known. They simulate each grain of the material explicitly, with all collisions fully resolved.

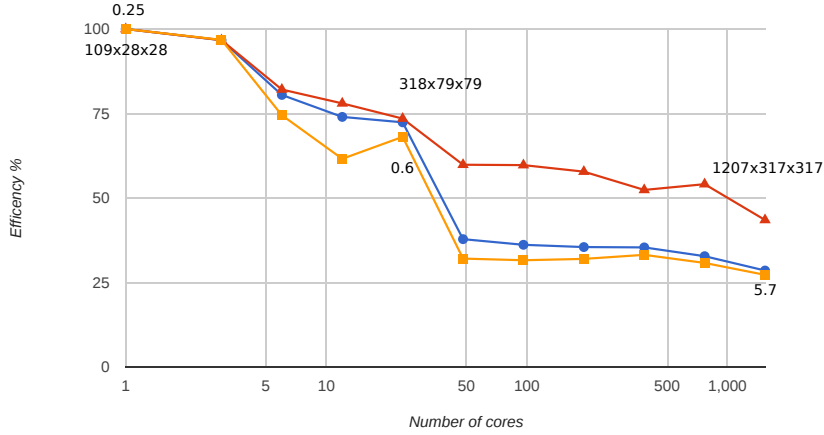

Figure 9: Parallel efficiency of the OpenFPM-based hybrid particle-mesh vortex code for a scaled-size problem (weak scaling). The problem size scales from $109 \times 28 \times 28$ mesh nodes on 1 processor core to $1207 \times 317 \times 317$ mesh nodes on 1536 cores (24 cores per node). We separately show the parallel efficiency for the PetSc Poisson solver (yellow squares), the OpenFPM parts of the code (red triangles) and the resulting overall scalability (blue circles). For three points, the problem sizes and the overall wall-clock time per time step in seconds are indicated next to the symbols. We note that the computational complexity of the Poisson solver is not linear with problem size.

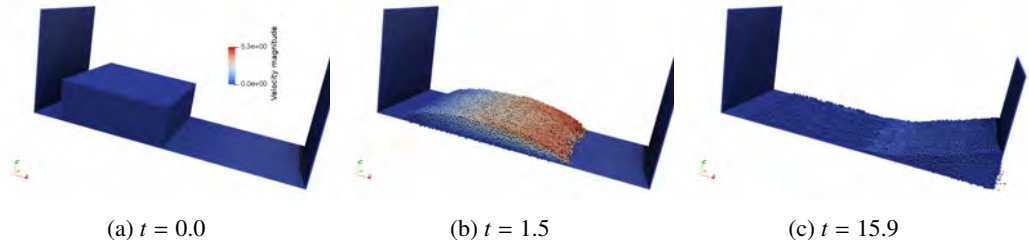

(a) $t = 0.0$   (b) $t = 1.5$   (c) $t = 15.9$

Figure 10: Visualization of the Discrete Element Method (DEM) simulation of an avalanche of spheres down an inclined plane (inclination angle: 30 degrees). OpenFPM output is shown for simulated times $t = 0.0$, $t = 1.5$, and $t = 15.9$. Particles are rendered as spheres, colored by velocity magnitude.

Force and torque balance over the grains then governs their Newtonian mechanics. The main difference to MD is that forces are only exerted by direct contact, and that contact sites experience elastic deformation. In order to correctly integrate these deformations over time, lists of contact sites between particles need to be managed. Since these lists are of varying length, both in time and space, and collisions involving ghost particles need to be properly accounted for in the lists of the respective source particles, parallelizing DEM is not trivial. Previously, DEM has been parallelized onto distributed-memory machines using the PPM Library [39], enabling DEM simulations of 122 million elastic spheres distributed over 192 processors [69].

We implement the same DEM simulation in OpenFPM in order to directly compare performance with the previous PPM implementation. We implement the classic Silbert grain model

**Algorithm 1** Vortex-in-Cell Method with two-stage Runge-Kutta (RK) time integration

1: **procedure** VORTEXMETHOD
2:     initialize the vortex ring on the mesh
3:     do a Helmholtz-Hodge projection to make the vorticity divergence-free
4:     initialize particles at the mesh nodes
5:     **while** $t < t_{\text{end}}$ **do**
6:         calculate velocity $u$ from the vorticity $\omega$ on the mesh (Poisson equation solver)
7:         calculate the right-hand side of Eq. 7 on the mesh and interpolate to particles
8:         interpolate velocity $u$ to particles
9:         *1st RK stage*: move particles according to the velocity; save old position in $x_{\text{old}}$
10:        interpolate vorticity $\omega$ from particles to mesh
11:        calculate velocity $u$ from the vorticity $\omega$ on the mesh (Poisson equation solver)
12:        calculate the right-hand side of Eq. 7 on the mesh and interpolate to particles
13:        interpolate velocity $u$ to particles
14:        *2nd RK stage*: move particles according to the velocity starting from $x_{\text{old}}$
15:        interpolate the vorticity $\omega$ from particles to mesh
16:        create new particles at mesh nodes (remeshing)

[70], including a Hertzian contact forces and elastic deformation of the grains, as previously considered [69]. All particles have the same radius $R$, mass $m$, and moment of inertia $I$. Each particle $p$ is represented by the location of its center of mass $r_p$. When two particles $p$ and $q$ are in contact with each other, the radial elastic contact deformation is given by:

$$\delta_{pq} = 2R - r_{pq}, \tag{9}$$

with $\boldsymbol{r}_{pq} = \boldsymbol{r}_p - \boldsymbol{r}_q$ the vector between the two particle centers and $r_{pq} = \|\boldsymbol{r}_{pq}\|_2$ its length. The evolution of the tangential elastic deformation $\boldsymbol{u}_{t_{pq}}$ is integrated over the time duration of a contact using the explicit Euler scheme as:

$$\boldsymbol{u}_{t_{pq}} = \boldsymbol{u}_{t_{pq}} + \boldsymbol{v}_{t_{pq}} \delta t, \tag{10}$$

where $\delta t$ is the simulation time step and $\boldsymbol{v}_{pq} = \boldsymbol{v}_{t_{pq}} + \boldsymbol{v}_{n_{pq}}$ are the tangential and radial components of the relative velocity between the two colliding particles, respectively. For each pair of particles that are in contact with each other, the normal and tangential forces are [70]:

$$\boldsymbol{F}_{n_{pq}} = \sqrt{\frac{\delta_{pq}}{2R}} \left( k_n \delta_{pq} \boldsymbol{n}_{pq} - \gamma_n m_{\text{eff}} \boldsymbol{v}_{n_{ij}} \right), \tag{11}$$

$$\boldsymbol{F}_{t_{pq}} = \sqrt{\frac{\delta_{pq}}{2R}} \left( -k_t \boldsymbol{u}_{t_{pq}} - \gamma_t m_{\text{eff}} \boldsymbol{v}_{t_{pq}} \right), \tag{12}$$

where $k_{n,t}$ are the elastic constants in normal and tangential direction, respectively, and $\gamma_{n,t}$ the corresponding friction constants. The effective collision mass is given by $m_{\text{eff}} = \frac{m}{2}$. In addition, the tangential deformation is rescaled to enforce Coulomb's law as described [70, 69]. The total resultant force $\boldsymbol{F}_p^{\text{tot}}$ and torque $\boldsymbol{T}_p^{\text{tot}}$ on particle $p$ are then computed by summing the contributions over all current collision partners $q$ and the gravitational force vector. We integrate the equations of motion using the second-order accurate leapfrog scheme, as:

$$v_p^{n+1} = v_p^n + \frac{\delta t}{m} F_p^{\text{tot}}, \qquad r_p^{n+1} = r_p^n + \delta t v_p^{n+1}, \qquad \omega_p^{n+1} = \omega_p^n + \frac{\delta t}{I} T_p^{\text{tot}}, \tag{13}$$

where $\boldsymbol{r}_p^n$, $\boldsymbol{v}_p^n$, and $\boldsymbol{\omega}_p^n$ are the center-of-mass position, velocity, and rotational/angular velocity of particle $p$ at time step $n$.

We simulate an avalanche down an inclined plane, which has previously been used as a benchmark case for distributed-memory parallel DEM simulations using the PPM Library [69]. The simulation, visualized in Fig. 10, uses 82,300 particles with $k_n = 7.849$, $k_t = 2.243$, $\gamma_n = 3.401$, $R = 0.06$, $m = 1.0$, and $I = 1.44 \cdot 10^{-3}$. The size of the simulation domain is $8.4 \times 3.0 \times 3.18$. Initially, all particles are placed on a Cartesian lattice inside a box of size $4.26 \times 3.06 \times 1.26$, as shown in Fig. 10a. The simulation box is inclined by 30 degrees by rotating the gravity vector accordingly and has fixed-boundary walls in *x*-direction, a free-space boundary in positive *z*-direction, and periodic boundaries in *y*-direction.

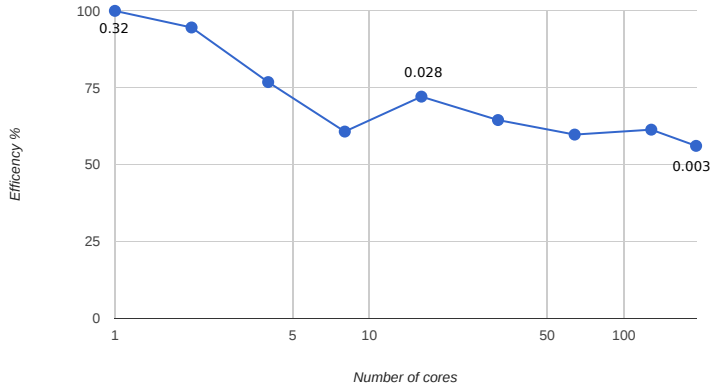

Figure 11: Strong scaling of the OpenFPM DEM simulation using a fixed problem size of $677,310$ particles distributed onto up to 192 cores using 8 cores on each cluster node. The numbers near the symbols indicate the absolute wall-clock time per time step in seconds.

We compare the performance of the OpenFPM DEM with the legacy PPM code [69] using the same test problem. In Fig. 11, we plot the parallel efficiency of the OpenFPM DEM simulation for a strong scaling on up to 192 processors. OpenFPM completes one time step with 677,310 particles on one core in 0.32 seconds, whereas the PPM-based code needs 1.0 second per time step for 635,780 particles. On 192 cores, OpenFPM completes a time step of the same problem in 3 ms with a parallel efficiency of 56%. In comparison, the PPM DEM client needs 11 ms per time step on 192 cores with a parallel efficiency of 47% [69]. This literature result is compatible with our present benchmark, as the PPM code was tested on a Cray XT-3 machine, whose AMD Opteron 2.6 GHz processors are about 3 times slower than the 2,5 GHz Intel Xeon E5-2680v3 of the present benchmark machine, indicating similar effective performance for both codes.

*4.6. Particle-swarm covariance-matrix-adaptation evolution strategy (PS-CMA-ES)*

One of the main advantages of OpenFPM over other simulation frameworks is that OpenFPM can transparently handle spaces of arbitrary dimension. This enables simulations in higher-dimensional spaces, such as the four-dimensional spaces used in lattice quantum chromodynamics [71, 72], and it also enables parallelization of non-simulation applications that require high-dimensional spaces, including image analysis algorithms [73] and Monte-Carlo sampling strategies [74].

A particular Monte-Carlo sampler used for stochastic real-valued optimization is the Covariance-Matrix-Adaptation Evolution Strategy (CMA-ES) [75, 76]. The goal is to find a (local) optimum of a (non-convex) function $f : \mathbb{R}^n \mapsto \mathbb{R}$. In practical applications, the dimensionality $n$ of the domain is 10 to 50. CMA-ES has previously been parallelized by running multiple instances concurrently that exchange information akin to a particle-swarm optimizer. The resulting particle-swarm CMA-ES (PS-CMA-ES) has been shown to outperform standard CMA-ES on multi-funnel functions [77], and an efficient Fortran implementation of it is available, pCMAlib [78].

Here, we implement PS-CMA-ES using OpenFPM in order to demonstrate how OpenFPM transparently handles high-dimensional spaces and also extends to non-simulation applications, such as sampling and computational optimization. In our implementation, each OpenFPM particle corresponds to one CMA-ES instance, hence implementing PS-CMA-ES through particle interactions across processors. To validate the OpenFPM implementation, we use the multi-modal test function $f_{15}$ from the IEEE CEC2005 set of standard optimization test functions [79]. In order to directly compare with pCMAlib, we limit the total number of function evaluations allowed to $5 \times 10^5$ and run both implementations 25 times each. We compare the success rate, i.e., the fraction of the 25 runs that found the true global optimum, and the success performance, i.e., the average best function value found across all 25 runs, in 10, 30, and 50 dimensions [77, 78]. The results from the OpenFPM-based implementation are identical with those from pCMAlib when using the same pseudo-random number sequence (not shown).

We also compare the runtime performance and parallel scalability of the OpenFPM-based implementation with the highly optimized Fortran pCMAlib. The results are shown in Fig. 12 for dimension 50. For dimensions 10 and 30, the results are analogous and not shown. Since the total number of function evaluations is kept constant at $5 \times 10^5$, irrespective of the number of cores used, this amounts to a strong scaling. However, the number of swarm particles is always chosen equal to the number of cores, as this is a hard requirement of pCMALib, while OpenFPM would not require this. In all cases, the OpenFPM implementation is about one third faster than pCMAlib.

Implementing arbitrary-dimensional codes using OpenFPM is straightforward, as the dimensionality is a template parameter in all data structures. While, of course, the memory requirement for mesh data structures grows exponentially with dimension, the size of particle data structures scales linearly. The code example in Listing 4.6 illustrates how the PS-CMA-ES data structures in 50 dimensions are defined in OpenFPM. All iterators and mappings work transparently. This example illustrates that OpenFPM naturally extends to problems in higher-dimensional spaces, which the original PPM Library [39] could not.

[Listing 4.6: OpenFPM code example for high-dimensional spaces]

```
1   constexpr int dim = 50;      // define the dimensionality
2
3   ///// Define the optimization domain as (-5:5)^dim
4   Box<dim,double> domain;
5   for (size_t i = 0; i < dim; i++) {
6           domain.setLow(i,-5.0);
7           domain.setHigh(i,5.0);
8   }
9
10  ///// Define periodic boundary conditions
11  size_t bc[dim];
12  for (size_t i = 0; i < dim; i++) {bc[i] = NON_PERIODIC;};
```

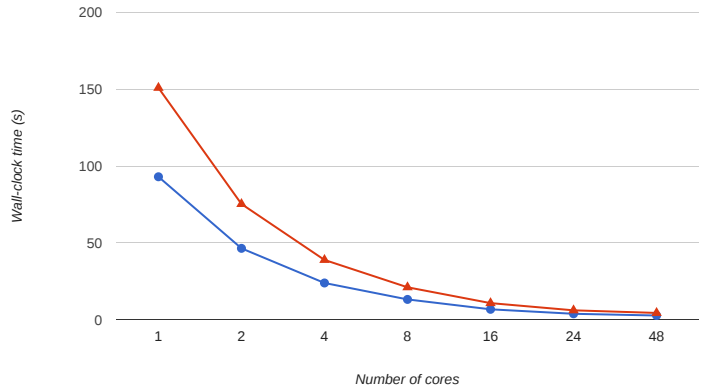

Figure 12: Strong scaling for the OpenFPM PS-CMA-ES client (blue circles) in comparison with the Fortran pCMALib (red triangles) scaling from 1 to 48 cores for IEEE CEC2005 test function $f_{15}$ in dimension 50. Shown is the minimum (over 25 independent repetitions) total wall-clock time in seconds for $5 \times 10^5$ function evaluations.

```
13
14   ///// There are no ghost layers needed for this problem
15   Ghost<dim,double> g(0.0);
16
17   ///// define the particles data structure
18   vector_dist<dim,double,aggregate<double,double[dim]>> particles(8,domain,bc,g);
19
20   ///// get an iterator over particles and loop over all of them
21   auto it = vd.getDomainIterator();
22   while (it.isNext()) {
23       .......    // do PS-CMA-ES here
24       ++it;
25   }
```

## 5. Conclusions

We have presented OpenFPM, an open-source framework for particle and particle-mesh codes on parallel computers. OpenFPM implements abstract data structures and operators for particles-only and hybrid particle-mesh methods [1]. The same abstractions were already implemented in the discontinued PPM Library [39], which has enabled particle-mesh simulations of unprecedented scalability and performance over the past 12 years [1]. OpenFPM extends this to a modern software-engineering framework using C++ Template Meta-Programming (TMP), continuous integration, and rigorous unit testing. OpenFPM provides a scalable infrastructure that allows implementing particle-mesh simulations of both discrete and continuous models, as well as non-simulation applications such as computational optimization [77] and image analysis [73]. The parallelization infrastructure provided by OpenFPM includes dynamic load (re-)balancing, parallel and distributed HDF5 file I/O, checkpoint-restart on different numbers of processors, transparent iterators for particles and mesh nodes, and adaptive domain decompositions. This infrastructure is supplemented with frequently used numerical solvers and a range of convenience functions, including direct VTK file output for visualization of simulation results using the open-source software Paraview [59].

We have described the architectural principles of OpenFPM and provided an overview of its functionality. We have then showcased and benchmarked the framework in six applications ranging from molecular dynamics simulations to 3D fluid mechanics to discrete element simulations, to optimization in high-dimensional spaces. Despite the automatic and transparent parallelization in OpenFPM, code performance and scalability in all examples was comparable to or better than those of state-of-the-art application-specific codes.

We have tested OpenFPM on up to 3072 processor cores, simulating systems with millions of degrees of freedom. For molecular dynamics, wall-clock times per time step were between 0.5 ms and 1 s, almost reaching the performance and scalability of the highly optimized LAMMPS code [28]. For SPH, OpenFPM outperforms the popular DualSPHysics CPU code [26] by about a factor of two, reaching GPU performance when using 270 CPU cores or more. Solving a finite-difference system on a regular Cartesian mesh, OpenFPM outperforms the highly optimized AMReX code [30] on small-scale problems, both in terms of scalability and performance. When using Vortex Methods [49] to simulate incompressible fluid flow, OpenFPM was able to compute vortex-ring dynamics at Re=3750 using 256 million particles on 3072 processors and achieved state-of-the-art parallel efficiencies in all benchmarks. Using DEM to simulate a granular avalanche down an inclined plane illustrated OpenFPM's capability to handle complex particle properties, such as time-varying contact lists, outperforming the previous PPM code [39, 69] by a small margin Finally, we illustrated the use of OpenFPM in high-dimensional problems by implementing PS-CMA-ES and comparing with the pCMAlib Fortran library [78]. This benchmark has shown the simplicity with which OpenFPM handles different space dimensions, while maintaining performance and scalability. Taken together, OpenFPM offers state-of-the-art performance and scalability at a reduced code development overhead. It overcomes the main limitations of the PPM Library [39] by extending to spaces of arbitrary dimension and allowing particles to carry arbitrary data types (C++ objects) as particle properties. It also adds automatic dynamic load (re-)balancing, transparent internal memory management and re-alignment, parallel checkpoint-restart, visualization file output, and custom distributed template expressions.

OpenFPM is going to be supported and developed in the long term. In the future, we plan to add the following functionalities to OpenFPM: transparent support for Discretization-Corrected Particle-Strength Exchange (DC-PSE) operators for the consistent discretization of differential operators on arbitrary particle distributions [37], an efficient distributed multi-grid solver for the general Poisson equation, 3D rendering capabilities for real-time *in-situ* visualization of a running simulation on screens and in virtual-reality environments, a compiler and development environment for application-specific language front-ends to OpenFPM, static (compile-time) and dynamic (runtime) code analysis [80] and optimization in order to reduce communication overhead to the required minimum, as well as support for adaptive-resolution particle representations [81, 45] and GPU calculations [82]. In addition, we will further improve performance and scalability, e.g., by optimizing the domain decomposition and sub-domain merging implementations and by using space-filling curves, such as Morton curves, to constrain processor assignment.

The source code of OpenFPM, virtual machines for various operating systems with a complete OpenFPM environment pre-installed, virtualized Docker containers, documentation, example applications, and tutorial videos are freely available from http://openfpm.mpi-cbg.de. We hope that the flexibility, free availability, performance, quality of documentation, and long-term support of OpenFPM will make it a standard platform for particles-only and hybrid particle-mesh simulations of discrete and continuous models on parallel computer hardware, as well as for non-simulation applications, such as evolutionary optimization strategies and particle-based image-analysis methods [83, 73].

### Acknowledgments

We thank all members of the MOSAIC Group for the many fruitful discussions. We particularly thank the early adopters and test users of OpenFPM whose feedback has helped improve the library throughout: Prof. Nikolaus Adams and Dr. Stefan Adami (both TU Munich, Germany), Prof. Marco Ellero (Swansea University, UK), Prof. Bernhard Peters (University of Luxembourg, Luxembourg), Prof. Bonnefoy (École des Mines Saint-Étienne, France), and Dr. Yaser Afshar (University of Michigan, Ann Arbor, USA). This project was supported in parts by the Deutsche Forschungsgemeinschaft (DFG) under project "OpenPME".